  \documentclass[12pt,preprint]{aastex}

\def\h2{H$_{2}$}

\def\n01{$N_{01}$} 
\def\t01{$T_{01}$}

\input epsf

\begin{document}

\title{THE FLUCTUATING INTERGALACTIC RADIATION FIELD AT REDSHIFTS 
Z = 2.3 -- 2.9 FROM HE~II AND H~I ABSORPTION TOWARD HE~2347-4342 } 

\author{J. MICHAEL SHULL\altaffilmark{1,2}, 
JASON TUMLINSON\altaffilmark{3}, 
MARK L. GIROUX\altaffilmark{4}, 
GERARD A. KRISS\altaffilmark{5}, 
and DIETER REIMERS\altaffilmark{6} }

\altaffiltext{1}{CASA, Department of Astrophysical and Planetary
     Sciences, University of Colorado, Boulder, CO 80309
     (mshull@casa.colorado.edu) }
\altaffiltext{2}{Also at JILA, University of Colorado and National
     Institute of Standards and Technology} 
\altaffiltext{3}{Department of Astronomy \& Astrophysics, University of
    Chicago, Chicago IL 60637 (tumlinso@oddjob.uchicago.edu)}
\altaffiltext{4}{Dept. of Physics and Astronomy, Box 70652, East 
     Tennessee State University, Johnson City, TN 37614 
     (giroux@polar.etsu.edu)}
\altaffiltext{5}{Space Telescope Science Institute, 3700 San Martin Dr.,
     Baltimore, MD 21218 (gak@stsci.edu)}  
\altaffiltext{6}{Hamburger Sternwarte, Universit\"at Hamburg, Gojenbergsweg 
     112, D-21029, Hamburg, Germany (dreimers@hs.uni-hamburg.de)} 

\begin{abstract}

We provide an in-depth analysis of the He~II and H~I absorption in the
intergalactic medium (IGM) at redshifts $z$ = 2.3--2.9 toward HE 2347-4342, 
using spectra from the {\it Far Ultraviolet Spectroscopic Explorer} (FUSE)
and the {\it Ultraviolet-Visual Echelle Spectrograph} (UVES) on the VLT
telescope.  Following up on our earlier study 
(Kriss et al.\ 2001, Science, 293, 1112), we focus here on two major topics:
(1) small-scale variability ($\Delta z \approx 10^{-3}$) in the ratio 
$\eta =$ N(He~II)/N(H~I); and (2) an observed correlation of high-$\eta$ 
absorbers (soft radiation fields) with voids in the (H~I) Ly$\alpha$ 
distribution.  These effects may reflect fluctuations in the ionizing sources
on scales of 1 Mpc, together with radiative transfer through a filamentary 
IGM whose opacity variations control the penetration of 1-5 ryd radiation
over 30--40 Mpc distances.   
Owing to photon statistics and backgrounds, we can measure optical depths 
over the ranges $0.1 < \tau_{\rm HeII} < 2.3$ and $0.02 < \tau_{\rm HI} < 3.9$,
and reliably determine values of $\eta \approx 4 \tau_{\rm HeII}/\tau_{\rm HI}$ 
over the range 0.1 to 460.  Values $\eta = 20-200$ are consistent with 
models of photoionization by quasars with observed spectral indices 
$\alpha_s = 0-3$.  Values $\eta > 200$ may require additional contributions 
from starburst galaxies, heavily filtered quasar radiation, or density variations. 
Regions with $\eta < 30$ may indicate the presence of local hard sources.  
We find that $\eta$ is higher in ``void'' regions, where H~I is weak or 
undetected and $\sim$80\% of the path length has $\eta > 100$. These voids 
may be ionized by local soft sources (dwarf starbursts) or by QSO radiation 
softened by escape from the AGN cores or transfer through the ``cosmic web". 
The apparent differences in ionizing spectra may help to explain the 
1.45 Gyr lag between the reionization epochs of H~I 
($z_{\rm HI} \sim 6.2 \pm 0.2$) and He~II ($z_{\rm HeII} \sim 2.8 \pm 0.2$).

\end{abstract}


\section{INTRODUCTION} 

In the last decade, studies of the intergalactic medium (IGM) have
been enhanced by high-resolution spectra of the Ly$\alpha$ absorption
lines of H~I and He~II.  This pioneering work includes spectrographic 
studies of H~I with the Keck telescope (Hu et al.\ 1995) and VLT (Kim, 
Cristiani, \& D'Odorico 2001) and of He~II by the {\it Hubble Space 
Telescope} (HST) (Jakobsen et al.\ 1994; Hogan, Anderson, \& Rugers 1997;
Heap et al. 2000; Smette et al.\ 2002) and the 
{\it Hopkins Ultraviolet Telescope} (HUT)
(Davidsen, Kriss, \& Zheng 1996).  With FUSE\footnote{This work is based 
on FUSE data obtained for the Guaranteed
Time Team by the NASA-CNES-CSA FUSE mission operated by the Johns Hopkins
University under NASA contract NAS5-32985. Observations with the NASA/ESA
Hubble Space telescope were supported by the Space Telescope Science
Institute, operated by AURA under NASA contract NAS5-26555. The VLT/UVES
observations were obtained at the Paranal Observatory of the European
Southern Observatory for program No.166.A-0106(A). }
we are now able to obtain moderate-resolution spectra of He~II absorption 
(Kriss et al.\ 2001, hereafter denoted KS01).  
Taken as a whole, these studies confirm the general theoretical picture 
(Cen et al.\ 1994; Zhang et al.\ 1995; Miralda-Escud\'e et al.\ 1996; 
Dav\'e et al.\ 2001) of the IGM as a fluctuating distribution of baryons, 
organized by the gravitational forces of dark matter and photoionized by 
high-redshift quasars and starburst galaxies (Haardt \& Madau 1996; 
Fardal, Giroux, \& Shull 1998, hereafter denoted FGS98; 
Madau, Haardt, \& Rees 1999).

Studies of absorption in the H~I and He~II Ly$\alpha$ lines at 
1215.670 \AA\ and 303.782~\AA\ respectively, are particularly important 
for understanding the transition of the high-redshift IGM from a neutral 
to ionized medium.  In practice, the He~II absorption becomes detectable 
in the far ultraviolet at redshifts $z \geq 2.1$ with FUSE, and at 
$z \geq 2.8$ with HST.  The ``reionization" epochs appear to occur at 
$z \approx 6.2\pm0.2$ for H~I (Becker et al.\ 2001; Djorgovski et al.\ 2001; 
Fan et al.\ 2002) and at $z \approx 2.8\pm0.2$ in He~II (Reimers et al.\
1997; KS01; Smette et al.\ 2002).  

The He~II Ly$\alpha$ absorption is generally much stronger than 
H~I Ly$\alpha$, by a factor denoted by the parameter 
$\eta = $N(He~II)/N(H~I), predicted theoretically (FGS98) to be 50--100 
for photoionization by quasar backgrounds.  For optically thin lines,
we can write $\eta \approx 4 \tau_{\rm HeII}/\tau_{\rm HI}$. 
The greater strength of He~II 
arises because it is harder to photoionize than H~I, owing to lower fluxes 
and cross sections at its ionizing threshold ($h\nu_T = 54.4$~eV), 
and because He~III recombines 5-6 times faster than H~II.  
Variations in $\eta$ can therefore be used to diagnose the sources 
and fluctuations of the metagalactic ionizing spectrum in both the 
continua of H~I (1 ryd) and He~II (4 ryd).  
The He~II absorption is also a good diagnostic of absorption
from low-density regions (``voids") in the IGM.  Because $\eta \gg 1$
and $\tau_{\rm HeII} = (\eta/4) \tau_{\rm HI}$, 
He~II can be tracked into much lower density regions than H~I. 
In principle, He~II absorption can be used to probe hotter, collisionally 
ionized gas at $T \approx 10^5$~K, produced by shock-heating 
(Cen \& Ostriker 1999;  Dav\'e et al.\ 2001).  

In our first paper (KS01), we analyzed FUSE observations of the 
fluctuating He~II absorption toward the bright (V = 16.1) quasar 
HE~2347-4342 at redshift $z_{\rm em} = 2.885 \pm 0.005$ (Reimers 
et al.\ 1997; Smette et al.\ 2002). 
This object was discovered in the Hamburg/ESO survey for bright QSOs
(Reimers, K\"ohler, \& Wisotzki 1996) and is one of the brightest
targets for FUV studies of He~II, with specific flux
$F_{\lambda} \geq 3 \times 10^{-15}$ erg~cm$^{-2}$ s$^{-1}$ \AA$^{-1}$
at 1180 \AA. The FUSE data include 376 ksec 
of observations between 1000--1187 \AA, taken at orbital night with a 
resolving power $R = (\lambda/\Delta \lambda) \approx 20,000$.  
Using the H~I Ly$\alpha$ data from Keck/HIRES, we analyzed both 
He~II and H~I Ly$\alpha$ forests over the interval $2.3 \leq z \leq 2.9$.  
About 50\% of the He~II features had H~I counterparts with column 
densities N$_{\rm HI} > 10^{12.3}$ cm$^{-2}$.  The He~II to H~I 
column density ratio $\eta$ ranged from 1--1000, with an average 
$\langle \eta \rangle \approx 80$.  

KS01 highlighted two major new results from the FUSE (He~II) data:
(1) the epoch of He~II reionization appears to occur in this sightline
 at $z_{\rm HeII} = 2.8 \pm 0.2$; (2) the range in the He~II/H~I ratio 
($\eta$) suggests
that the ionizing background is highly variable throughout the IGM.
Recent HST observations (Telfer et al.\ 2002) show a broad distribution 
in QSO spectral indices, from $\alpha_s \approx 0 - 3$.
Over the observed redshift interval, $2.3 < z < 2.7$, the IGM toward
HE~2347-4342 is ``translucent" in He~II, with optical depths
fluctuating between $\tau_{\rm HeII} \approx 0.1$ and $\tau_{\rm HeII} > 3$.
Above $z > 2.72$, the IGM becomes relatively opaque in He~II, except for
three narrow transmission windows around 1160~\AA\ ($z = 2.82$) and 1175~\AA\
($z = 2.87$) noted previously (Smette et al.\ 2002; KS01).  Below $z < 2.72$,
the general trend is that the IGM becomes more transparent,
with $\langle \tau_{\rm HeII} \rangle \approx 1$, slowly decreasing at
lower redshift.  

Except for the high-opacity regions ($z = 2.7-2.9$), the observed 
He~II optical depths and He~II/H~I ratios are in reasonable agreement 
with models of IGM photoionization by a QSO-type power-law spectrum,
$F_{\nu} \propto \nu^{-\alpha_s}$, with $\langle \alpha_s \rangle =
1.8 \pm 0.2$ (Zheng et al.\ 1997; Telfer et al.\ 2002).  Models of
the He~II/H~I ratios in the IGM produced by QSO sources, filtered and
reprocessed by the IGM, have been presented by FGS98 and Madau, Haardt, 
\& Rees (1999). The high values ($\eta > 200$) suggest additional
contributions from softer ionizing sources (starburst galaxies) or
heavily-filtered quasar radiation.

In this second paper on HE~2347-4342, we provide an in-depth analysis  
and scientific discussion of these same FUSE He~II data, in combination 
with new optical H~I data taken with the VLT/UVES spectrograph (D'Odorico 
et al.\ 2000). We describe here the characterization and subtraction of 
the background in the LiF (1000--1187 \AA) channels of FUSE.
An independent analysis of both the LiF and SiC channels (920--1000~\AA)
is described in a separate paper (Zheng et al.\ 2003).  
We compare FUSE and HST/STIS data over the common interval from 
1150--1187 \AA.  We then analyze the
He~II and H~I Ly$\alpha$ absorbers, characterize their distributions in 
redshift and column density, and derive optical depths, $\tau_{\rm HeII}$ 
and $\tau_{\rm HI}$.  The observed absorption ratios, 
$\eta \approx 4\tau_{\rm HeII}/\tau_{\rm HI}$, can be 
used to understand the photoionizing spectra of the sources of the 
metagalactic radiation.  We also explore the nature of the 
``void He~II absorption", for which He~II is an especially good probe. 
Finally, we address some of the open issues about the high-$z$ IGM, 
including the He~II reionization epoch, the ionization history of the 
IGM, and the range of ionizing spectra in the gaseous voids and filaments.  

The remainder of this paper is organized as follows.  In \S~2, we give 
details of our new FUSE data extraction and background subtraction. 
With the VLT data, we then derive the relevant parameters that characterize 
the two Ly$\alpha$ forests ($\tau_{\rm HI}$, $\tau_{\rm HeII}$, $\eta$). 
In \S~3 we discuss the scientific implications of these observations
for the IGM and its ionization state at $z < 2.9$, just after
the apparent He~II reionization epoch. We conclude in \S~4 with a 
summary of the highlights of this new work and suggestions for
future studies of He~II with FUSE and HST.

\section{DATA ANALYSIS} 

\subsection{Provenance of the data}

As described in KS01, HE~2347-4342 was observed with FUSE in two separate
campaigns.  The first observation, from 17 to 27 August 2000, comprised
351,672 s, 192,610 s during orbital night.  The second campaign, running
from 11 to 21 October 2000, accumulated 249,717 s, with 183,630~s during
orbital night.  For each observing campaign, HE~2347-4342 was centered
in the $30'' \times 30''$ apertures. To maintain optical alignment of the
four channels during these campaigns, FUSE was offset every other day to
the nearby UV-bright star WD2331-475 to perform routine adjustments to
the mirror positions.  FUSE has an effective resolution $R \approx 20,000$
(Moos et al.\  2000), which corresponds to a wavelength resolution element
$\Delta \lambda = 0.05$~\AA\ at 1000 \AA. To avoid uncertainties in the
effective spectrograph resolution and because the resolution element is
highly oversampled, we bin our data to at least 0.05 \AA.  This is 
approximately the minimum linewidth of the H~I features we examine.

We chose to neglect the wavelength range provided by the FUSE SiC
detectors for two reasons. First, the SiC channels have a throughput
that is $\sim$3 times less than that of the LiF channels ($\lambda >
1000$~\AA). Second, the higher effective area SiC channels (1a and 1b)
appear near the bottom edge of the detectors, so that it is impossible
to obtain a background spectrum based on unilluminated regions that
bracket the spectrum. Recently, Zheng et al. (2003) have succeeded in
extracting a useful spectrum from the SiC channels by performing careful
burst removal, pulse-height screening, and background modeling.  

HE~2347-4342 was observed during four nights (October 7-9 and
October 19, 2000) using the UVES installed at the second VLT
Unit Telescope (Kueyen).  Twelve exposures were made in the
dichroic mode using standard settings for central wavelengths
of 3460/4370 \AA\ in the blue and 5800/8600 \AA\ in the red.
The integration time for each wavelength range was 6 $\times$
60 min.  With a slit width of 1 arcsec, the resulting spectral
resolution is about 44,000. The raw frames were reduced at
Quality Control Garching, using the UVES pipeline data reduction
software (Ballester et al.\ 2000). Finally, the individual
vacuum-barycentric corrected spectra were coadded, resulting in
an effective S/N ratio of roughly 100 per 0.05~\AA\ bin. 
When compared to the FUSE He II data, the H~I spectrum is resampled 
to the same redshift vector and bin size; these sizes vary
in our analysis from 0.05 to 0.2~\AA\ (see \S~2.3). 

\subsection{Data Calibration and Background Subtraction}

The faint flux levels of our target present unusual challenges to
the instrumental calibration of FUSE. We can simply apply stock flux
corrections and wavelength solutions to our data with high confidence
that these solutions are stable in time and insensitive to the source
flux. However, the detector background correction that is usually applied
to FUSE data by the {\sc CALFUSE} pipeline (up to the latest version
2.2.1 as of August 2002) is insufficient to model correctly the spatial
variations in the background for a source flux at $(1-3) \times  10^{-15}$ erg
cm$^{-2}$ s$^{-1}$ \AA$^{-1}$.  In our 376 ksec of data, we collected
roughly 5 - 20 source counts per 0.0067 \AA\ column in the extracted
source spectrum, spread over 50 - 100 rows in the 2D detector image. This
faint source level distributed over a large detector area leaves 1-2
source + background counts per element in the illuminated regions of
the detector image. This small source flux makes the background
subtraction an unusually crucial piece of the overall calibration.

At the faint flux levels presented by HE~2347-4342, an accurate
extraction of a background-subtracted, calibrated source spectrum
from the two-dimensional recorded data required customized processing.
We began our analysis with time-tagged photon lists representing all the
data. We use only the nighttime data to avoid difficult problems related
to scattered airglow light on the detectors.  Up to the background
subtraction, our analysis used the standard {\sc CALFUSE} pipeline
(v2.2.1) to process the photon lists, screen for and remove detector
bursts, and create a 2D detector image.  We applied the default CALFUSE
day/night screening and truncated the range of detector pulse heights
from channels 4 - 16 to minimize backgrounds.  The resulting detector
image has 16,384 ``columns'' (the $x$ direction) and 1024 ``rows'' (the
$y$ direction). These elements represent the digital quantization of
the analog positions of photons striking the photocathode in the FUSE
microchannel plate detectors (Sahnow et al. 2000). In this analysis, 
we avoid the term ``pixels'' because FUSE quantizes an analog quantity
instead of using literal ``pixels'' as in CCDs.  At this step, we depart
from the CALFUSE procedure to produce a customized background subtraction
with the scheme described below.

Prior to constructing background vectors, for each channel/segment
pair (LiF1a, LiF1b, LiF2a, LiF2b), we added all the columns by summing
in $x$. This summation reveals the $y$ positions of the source-illuminated
regions, including residual airglow lines present during orbital
night.  It establishes the height of the rectangular extraction
windows and the $y$ limits of the unilluminated background regions (see
Table 1).  We derive an $x$-dependent background correction by taking
unilluminated regions of the detector, above and below the illuminated
source plus background region, and extracting a 1D background spectrum as
a function of $x$. This scheme gives us detailed information about the
large spatial $x$ variations in the background, but requires averaging
over the variations in $y$ that may occur beneath the illuminated portion.

To construct the 1D background spectrum for an extraction window
of height $n$ elements, we randomly drew $n$ elements (at each $x$)
from the background windows and summed over the number of elements
in the extraction window. This sampling is done 300 times, and the
final background spectrum is the average of the 300 trials at each $x$
position. For example, for LiF1a we randomly drew 80 bins from the 150
bins in the three available background windows. This sampling is done
300 times and the final background and uncertainty are drawn from the
average and 1$\sigma$ variation in the distribution of samples. The final
background vector approximates the number of background counts collected
in each row over the height of the extraction window. The random sampling
accounts for fluctuations with $y$ and assigns an error bar to
the background subtraction, which is propagated through our analysis. The
final calibration steps apply the CALFUSE wavelength and flux solutions. \\ 

For the FUSE data, we adopt the continuum determined from
STIS/G140L data by KS01, adjusted to account for the contributions
(at the 5\% level) of an intervening Lyman-limit system at $z = 2.739$ 
with N(H~I) $= 2.5 \times 10^{17}$ cm$^{-2}$.  This LL system makes
a 5\% contribution to the absorption at 1100 \AA.  
The extrapolation gives close agreement in He~II transmission
in the region (1157--1180~\AA) where our data and the STIS results 
(Smette et al.\ 2002) overlap.  The full recovery of He~II to this 
continuum in transmission windows near 1000 \AA\ gives us confidence 
that this extrapolation is acceptable at these wavelengths.

The observations reported here push the limits of FUSE's design
capabilities and exceed the limitations of the {\sc CALFUSE} automatic
data-reduction software. Our custom reduction was made
as simple and accurate as possible, with careful attention to systematic
errors and stability of the background subtraction.  It appears that
the differences between the KS01 reduction and the current work are
comparable to the differences between overlapping segments in our new
extraction, which suggests that possible systematic errors associated
with our background subtraction method and choice of extraction windows
are well controlled.  Furthermore, the differences between the October
2000 and August 2000 data and between LiF1 and LiF2 are consistent within 
statistical errors, and are not sensitive to the choices made
in constructing the background spectrum.  Table 2 shows the mean He~II
transmission over several wavelength bands, the results of a cross-check
with earlier studies by STIS (Smette et al.\ 2002) and FUSE (KS01).

Figure 1 shows an overlay of the FUSE (He~II) data with the VLT/UVES
(H~I) data.  The H~I wavelength scale has been divided by a factor of
4.00178 (the ratio of He~II and H~I rydberg constants is not precisely 4.0) 
to align the H~I and He~II absorbers.  One clearly sees the much
larger He~II absorption ($\eta \gg 1$) predicted by theory.  Other notable
features include the transmission windows around
1132~\AA, 1160~\AA, and 1175~\AA, in close proximity to regions of high
He~II opacity.  We also note the frequent voids in H~I absorption.  

Figure 2 shows the broad-band transmission in the He~II Ly$\alpha$
forest from 1000--1180~\AA.  As noted by KS01,
this shows the expected increase in broadband transmission towards
shorter wavelengths.  However, on a fine-grained scale, $\Delta \lambda
= 0.05-0.20$~\AA, the He~II opacity shows substantial variations.  In our
analysis, we use various bin sizes to explore the dependence of H I
and He II opacity on resolution. The smallest bin size we use
is 0.05 \AA, or $\Delta z = 1.6 \times 10^{-4}$, which corresponds to 
the resolution element of FUSE.

\subsection{The He~II/H~I Ratio ($\eta$)}
 
As defined, $\eta$ expresses the ratio of He~II to H~I column densities
and can be used to infer properties of the ionizing radiation field.  
Owing to photon statistics and backgrounds, we believe that we can measure
optical depths over the ranges $0.1 < \tau_{\rm HeII} < 2.3$ in the FUSE
data and $0.02 < \tau_{\rm HI} < 3.91$ with the VLT/UVES.
These values correspond to 10\% uncertainties (FUSE) and 2\% uncertainties
(VLT/UVES) in measuring the optical depths of either strong or weak lines.     
In this paper, unlike KS01, we do not make the assumption that He~II 
absorption occurs in distinct lines. Instead of finding 
$\eta =$N(He~II)/N(H~I) from the column density ratio, we use the observed 
opacities per resolution element to define
 $\eta = 4 \tau_{\rm HeII} / \tau_{\rm HI}$ (the extra factor of 4 arises
from bandwidth).  As a consequence of these ranges for measuring optical 
depths, we can reliably determine values of $\eta$ between 
approximately 0.1 and 460.

Figure 3 shows the observed distribution of $\eta$ with redshift, from
$z = 2.30 - 2.90$, for two different wavelength bin sizes (0.05~\AA\
and 0.20~\AA). We also distinguish the absorbers in``voids" 
($\tau_{\rm HI} < 0.05$) from those in the H~I ``filaments" (stronger 
absorbers with $\tau_{\rm HI} > 0.05$).  Note that $\eta$ is
systematically higher in the voids.  Binning the data from 0.05~\AA\
 to 0.20 \AA\ reduces the number of points in the low ($\eta = 0.1-10$) 
and high ($\eta = 200-400$) ends of the range, but it does not change
the overall distribution. 

In practice, we must censor the data to avoid unphysical values of $\eta$.  
First, we exclude points with $F_{\lambda} < 0$. We also excise points 
within 1$\sigma$ of either the zero level or the continuum, on the
grounds that they are indistinguishable statistically from 0 and 1,
respectively. Thus, the $\eta$ distributions shown below represent only
a subset of the complete dataset.  With these criteria for censorship,
we find $\langle \eta \rangle \approx 45$.  If the FUSE points
within 1 $\sigma$ of zero are used to provide lower limits on
the He II optical depth, and thus lower limits on $\eta$, a Kaplan-Meier
survival statistics calculation implies $\langle \eta \rangle \approx 110$.
However, this neglects the upper limits available for $\eta$.

The uncensored mean value, $\langle \eta \rangle = 80$, is in good
agreement with theoretical predictions (FGS98) for an ionizing background
produced by QSOs with mean spectral index $\langle \alpha_s \rangle = 1.8$
(Zheng et al.\ 1997; Telfer et al.\ 2002).  The wide range in $\eta$,
as well as the small-scale variations in redshift, were attributed in
KS01 to the possible contributions of hard and soft sources
to the ionizing background.  However, this assertion deserves further
scrutiny with a physical model of such sources and their spatial frequency.

In photoionization equilibrium, the He~II/H~I ratio can be expressed 
(FGS98) as: 
\begin{equation}
   \eta = \frac {n_{\rm HeIII}} {n_{\rm HII}} 
          \frac {\alpha_{\rm HeII}^{(A)}} {\alpha_{\rm HI}^{(A)}}  
          \frac {\Gamma_{\rm HI}} {\Gamma_{\rm HeII}} 
   \approx (1.70) \frac {J_{\rm HI}} {J_{\rm HeII}} 
        \frac {(3 + \alpha_4)}{(3 + \alpha_1)} T_{4.3}^{0.055} \; . 
\end{equation}
Here, $\alpha_{\rm HI}^{(A)}$, $\alpha_{\rm HeII}^{(A)}$, 
$\Gamma_{\rm HI}$, and $\Gamma_{\rm HeII}$ are the case-A recombination 
rate coefficients and photoionization rates for H~I and He~II, and  
$J_{\rm HI}$ and $J_{\rm HeII}$ are the specific intensities
of the radiation field at the H~I (912~\AA) and He~II (228~\AA) edges. 
The parameters $\alpha_1$ and $\alpha_4$ are the local spectral indices 
of the ionizing background at 1 and 4 ryd, respectively, which 
provide minor corrections to the photoionization rates.
We adopt case-A hydrogenic recombination rates to H~I and He~II, 
appropriate for the Ly$\alpha$ forest absorbers with very low neutral 
fractions. Over the temperature range $16,000~{\rm K} < T < 32,000$~K, 
we approximate
$\alpha_{\rm HeII}^{(A)} = (1.36 \times 10^{-12}~{\rm cm}^3~{\rm s}^{-1})
T_{4.3}^{-0.700}$ and
$\alpha_{\rm HI}^{(A)} = (2.51 \times 10^{-13}~{\rm cm}^3~{\rm s}^{-1})
T_{4.3}^{-0.755}$, where $T_{4.3} \equiv (T/10^{4.3}~{\rm K})$.  
We also assume that $n_{\rm He}/n_{\rm H} = 0.0789$ ($Y = 0.24$) and 
that H and He are predominantly fully ionized ($n_e = 1.16 n_H$).

The metagalactic ionizing background, $J_{\nu}(z)$, can be modeled
(Shull et al.\ 1999) from the intrinsic spectra of the ionizing sources 
(spectral index $\alpha_s$) filtered through the IGM.  Cosmological
radiative transfer softens the spectrum and produces a steeper 
effective index, $\alpha_{\rm eff} > \alpha_s$.  For example, 
one can write the ratio of specific intensities at 1 and 4 ryd as 
$J_{\rm HI}/J_{\rm HeII} = 4^{\alpha_{\rm eff}}$, where $\alpha_{\rm eff}
\approx 2.8$ for $\alpha_s = 2.0$ (see Fig.\ 10 of FGS98).  From
equation (1), we then find $\langle \eta \rangle \approx 80$ for
$T_{4.3} \approx 1$ and $\alpha_1 \approx \alpha_4$.   

Figure 4 shows the observed range in ionizing spectral indices, 
$\alpha_{\rm EUV}$, of radio-quiet QSOs from the HST spectral survey 
(Telfer et al.\ 2002). This survey favors blue quasars with fairly clean 
lines of sight. This might bias the mean of the $\alpha_{\rm EUV}$ 
distribution to be bluer, and select against ones with very red
$\alpha_{\rm EUV}$. However, both radio-loud and radio-quiet samples show
the same broad distribution of spectral indices.  
The upper curves in Fig.\ 4 show the spectral indices,
$\alpha_s$, required to reproduce the observed values of $\eta$, using 
equation (1).  We use bin-by-bin 
data for optical depths and convert them to the required
$\alpha_s$, assuming $T_{4.3} = 1$ and $\alpha_1 = \alpha_4$. In this
approximation, $J_{\rm HI}/J_{\rm HeII} = 4^{\alpha_s}$, and
$\eta = 1.70 \times 4^{\alpha_s}$ from equation (1).
It is also possible that some He~II/H~I variations could be produced by
hot shocked gas ($10^{5-6}$~K), for which $\eta$ ranges from 10--500 in
pure collisional ionization equilibrium (Sutherland \& Dopita 1993).    

We note that the range of $\alpha_s$ (Fig.\ 4) is considerably larger than 
that of $\alpha_{\rm EUV}$, suggesting additional spectral softening. The 
absorbers in voids have higher spectral indices ($\langle \alpha_s 
\rangle \approx 3.0$) than those in filaments ($\langle \alpha_s 
\rangle \approx 1.2$), suggesting that the void absorbers 
may experience an ionizing radiation field softer than that of
the ``bare QSOs". However, since $\eta$ has fine-grained variation over
$\Delta z \approx 10^{-3}$, it may be inappropriate to use a mean 
metagalactic radiation field for all absorbers in the IGM.   
The ratio, $J_{\rm HI}/J_{\rm HeII}$, must fluctuate due to ``local effects", 
and the observed wide range of QSO spectral indices is proposed to account 
for most of these variations.  
This poses a paradox: the space density of AGN at $z \approx 2.5$ 
is too sparse (mean distance of 30 Mpc between QSOs) to produce the 
observed $\eta$-fluctuations ($\Delta z \approx 10^{-3}$ or 
$1.3 h_{70}^{-1}$~Mpc at $z \approx 2.5$).  We return to this issue in \S~2.4. 

\subsection{Statistics and Distribution of He~II/H~I Ratios}

In this subsection, we explore the variations of $\eta$ in greater detail,
using an analysis of their correlations with voids and filaments in the
Ly$\alpha$ forest.  The large number of $\eta < 30$ points is unexpected,
as it implies that there is a substantial volume in the universe
where the extragalactic radiation background has a very hard spectrum.
With a series of Monte Carlo calculations, we now explore the degree to which
the high fraction of these points is due to bias in the points chosen to
calculate $\eta$.  We begin with the H~I spectrum and generate
a new set of flux points based on the uncertainties given in that spectrum.
We then convert this new set of flux points into an H~I optical depth,
$\tau_{\rm HI}$.  With the assumption that the underlying value of $\eta$ 
is actually a constant $\eta_c$, we then generate a corresponding set of 
He~II optical depths, convert these optical depths into a set of flux points, 
and vary these points according to the uncertainty vector in our FUSE spectrum.  
We analyze this simulated spectrum in the same way as outlined for the 
observed spectrum, excising points with $F_\lambda < 0$, and within
1~$\sigma$ of zero or the continuum.  Finally, we calculate a new
set of values for $\eta$.

After performing a set of 1000 simulations, we arrive at the
following conclusions:
(1) For any assumed values of $\eta_c$, there is a trend to
    produce a simulated set of points with $\langle \eta \rangle < \eta_c$;
(2) Nevertheless, even for a value of $\eta_c$ chosen so that 
    $\langle \eta \rangle$ is the observed censored value (specifically,
    $\eta_c = 52$ and $\langle \eta \rangle \approx 45$ as in \S~2.3), the 
    fraction of low-$\eta$ flux points in the data is larger than that in 
    the simulations.  In the data, 55\% of all points have $\eta < 30$,
    and 30\% of all points have $\eta < 10$. The mean fractions of such 
    points in the simulations are 27\% and 14\%, respectively.
    No simulation had as many low-$\eta$ points as the data. 

Figure 3 indicates another check of the reality of the low $\eta$ points.  
We have made the same observational cuts in $\tau_{\rm HI}$ for the points 
in the upper left and lower left plots.  In the lower left panel, where the 
bin size is quadrupled and the S/N approximately doubled, the distribution 
in $\eta$ is much less likely to be affected by statistical fluctuations in 
the measured flux.  The persistence of numerous low-$\eta$ points is 
reassuring to the interpretation that these points are real.  
As we will show, we believe the fine-scaled $\eta$ variations are 
statistically significant and arise from a combination of intrinsic 
variation in AGN spectral indices, coupled with radiative transfer effects 
in a non-uniform IGM.

Although our analysis assumes that all absorption in the VLT spectrum is 
due to H~I, and that all absorption in the FUSE spectrum is due to He~II,
both spectra are subject to metal-line contamination. Songaila (1998) identified 
metal-line systems (C~IV and Si~IV) at $z=2.43814$ and $z=2.63449$, as well 
as metal lines connected with a Lyman-limit (LL) system (Reimers et al.\ 1997; 
Smette et al.\ 2002).  The LL system has hydrogen continuum optical
depth $\tau_{\rm cont} = 1.6$ at $z_{\rm LL} = 2.739$, corresponding to 
N(H~I) = $2.5 \times 10^{17}$ cm$^{-2}$ and can be seen in
Figure 1 as the two saturated H~I absorbers labeled on the He~II scale at 
1135 and 1136 \AA\ (the corresponding He~II lines fall within a detector gap, 
and so are not visible).  In addition, the lack of a proximity effect in the 
He~II Ly$\alpha$ spectrum toward HE~2347-4342 implies a column density
N(He~II) = $2.4\times 10^{18}$ cm$^{-2}$ at $z_{abs} \approx z_{em}$
(Reimers et al.\ 1997; Smette et al.\ 2002) [see \S~3],
which has highly ionized metal lines associated with it.  Smette et al.\ 
(2002) report components with log N(C~IV) = 12.8 and 12.1 as well as 
log N(O~VI) = 13.6 associated with this absorption system.  In general, 
for a range of assumed properties of this gas, the possible corresponding 
transitions are likely to be quite weak in the FUSE or VLT spectra.

There also is the possibility that metal lines (C~IV, Si~III) associated 
with lower redshift Ly$\alpha$ absorbers may be present in portions
of the VLT spectra.  This is unlikely to impact our statistical conclusions 
strongly.  At higher redshift, on average,
$\langle {\rm N(C~IV)/N(H~I)} \rangle = 3\times 10^{-3}$ for 
N(H~I) $\approx 10^{14.5}$ cm$^{-2}$.  This would imply an equivalent width 
of only 1.2~m\AA\ for N(H~I) $\approx 10^{14.0}$ cm$^{-2}$, and much less
for the more common lower column density absorbers.  Finally, there is an 
unknown amount of H$_2$ Lyman and Werner band absorption ($\lambda < 1120$~\AA).
This is a sightline with relatively low extinction, so that heavy Galactic 
H$_2$ contamination is unlikely.  We have explored the extent to which 
metal-line contamination would bias our statistical results. Simulating 
the effects of contamination, we randomize 10\% and 20\% of the opacity 
bins and reanalyze the data.  In 100 realizations of each process,
we have found that our statistical conclusions are robust.

Next, we have examined the distribution of $\eta$, both in regions of
high H~I opacity (denoted ``filaments") and in low-opacity regions
(denoted ``voids").  Figure 5 shows the cumulative distributions
of $\eta$ in five different regions, two voids and three filaments,
all in the redshift range $2.35 < z < 2.44$. One can immediately see that 
the voids contain many more absorbers with large values of $\eta$
(soft ionizing sources) than the filaments.  This effect was also noted 
in Figure 4.   The radiation field in the voids could be dominated by 
starburst galaxies, whose massive stars lack the strong He~II continua 
of QSOs.  Alternatively, the AGN radiation reaching the voids may have 
been more strongly filtered by passage through dense filaments in the
``cosmic web".   

Figure 6 shows the distribution of optical depths, $\tau_{\rm HeII}$, 
in two redshift intervals.  In each panel, we show distributions for 
``voids" ($\tau_{\rm HI} < 0.05$) and ``filaments" ($\tau_{\rm HI} > 0.05$) 
in the H~I Ly$\alpha$ forest.  The flat distributions in this logarithmic 
plot between $0.1 < \tau_{\rm HeII} < 1$ are consistent with 
$N(\tau_{\rm HeII}) \propto \tau_{\rm HeII}^{-1}$, steepening at 
$\tau_{\rm HeII} > 1$.  This distribution is somewhat flatter than the 
corresponding H~I column density distribution, 
$f(N_{\rm HI}) \propto N_{\rm HI}^{-\beta}$, with
$\beta = 1.4-1.5$ for $1.5 < z < 4$ (Kim et al.\ 2001).

Figure 7 shows a closeup of one of the ``void" regions ($z = 2.40-2.42$)
in which the He~II transmission is unusually large, again with many 
high-$\eta$ absorbers suggestive of local soft sources.  However, the 
difficulty with this simple explanation is that the expected high space density   
of (dwarf starburst) galaxy sources should produce a uniform ionizing
background, not the observed fluctuating field of $\eta$
values.  An alternative interpretation is that the hard-spectrum QSOs 
might preferentially be located in the overdense regions associated with
H~I filaments.  In this case, the filament absorbers would see a
harder spectrum (lower $\alpha_s$, lower $\eta$) than the
voids.  Ultimately, these speculations need to be confirmed by 
cosmological radiative transfer calculations in an inhomogeneous IGM.

\section{DISCUSSION}
    
In this paper, we have used FUSE (He~II) and VLT/UVES (H~I) data to analyze 
the fluctuating intergalactic absorption between $z = 2.3-2.9$ toward the 
high-redshift QSO HE~2347-4342.  This interval appears to span the He~II 
reionization epoch at $z_{\rm HeII} = 2.8 \pm 0.2$, which is considerably 
lower than that, $z_{\rm HI} = 6.2 \pm 0.2$, estimated for hydrogen 
reionization.  For a standard WMAP cosmology (Spergel et al.\ 2003), with 
$H_0 = 71 \pm 4$ km~s$^{-1}$ Mpc$^{-1}$, $\Omega_m = 0.27 \pm 0.02$, 
and $\Omega_{\Lambda} = 0.73 \pm 0.02$, the interval of time between 
these redshifts corresponds to $\Delta t = (1.45 \pm 0.13) h_{70}^{-1}$ Gyr.

As in KS01, we confirm the slow reduction of $\tau_{\rm HeII}$ at lower 
redshifts, $z < 2.7$, in good agreement with theoretical expectations (FGS98)
of an IGM photoionized by QSOs.  The He~II ``reionization epoch" in this one 
sightline is characterized by several regions of high optical depth 
between $z \approx 2.75-2.85$, separated by narrow transmission 
windows at $z = 2.82$ and 2.87.  Clearly, FUSE and/or HST studies of 
additional sightlines are needed to understand the cosmic variance in 
these results. 

\noindent 
Our major findings on IGM structure and sources of ionizing 
intergalactic radiation are: 
\begin{enumerate}

\item We see fine-grained variations in He~II/H~I ratios ($\eta$) 
   on scales $\Delta z \approx 10^{-3}$, corresponding to physical
   lengths $\sim 1.3 h_{70}^{-1}$ Mpc at $z \approx 2.5$.

\item The $\eta$-variations appear to reflect ``local radiation effects" 
   from QSOs whose spectral indices range from $\alpha_s = 0-3$.  A few 
   regions with $\eta < 30$ could lie close to hard ionizing sources.

\item  The He~II/H~I ratio appears to be larger in voids in the H~I Ly$\alpha$
   forest ($\tau_{H I} < 0.05$) compared to filaments.  In these voids, up to 
   80\% of the pathlength has $\eta > 100$.

\end{enumerate}

\noindent
These results suggest that, if both He~II and H~I are photoionized,
the metagalactic radiation field is highly variable and strongly affected
by local sources.  In principle, the $\eta$-fluctuations 
could arise solely from variations in IGM opacity. But, in view of the 
broad observed range in AGN spectral indices, $\alpha_s = 0-3$, it seems
probable that the fluctuations arise from spatial variations in the
ionizing fluxes (``local effects").   Moreover, observations of the 
Ly$\alpha$ absorbers at $z = 1-3$ (Kim et al.\ 2001) suggest a low spatial
frequency of strong absorbers that dominate the opacity at the 
H~I or He~II ionization edges.  Optical depths unity in these 
continua correspond to H~I column densities $10^{17.2}$ cm$^{-2}$ and 
$10^{15.8} (100/\eta)$ cm$^{-2}$, respectively.  Such systems are too rare 
in the Ly$\alpha$ forest at $z \approx 2.5$ to produce the fine-grained 
variations over $\Delta z \approx 10^{-3}$.

However, the simple explanation that QSO emissivity variations translate
into fluctuations in $\eta$ needs a more firm physical basis.  The 
variable-index QSOs at $z = 2.3-2.8$ must produce spatial variations 
in the radiation field over Mpc-scales ($\Delta z = 10^{-3}$), whereas  
the co-moving space density of bright ($L^*$) QSOs at $z = 2.5$ is 
$\phi_{\rm QSO} \approx 10^{-6}~h_{70}^3$ Mpc$^{-3}$, which corresponds
to mean physical separation, 
$\langle d_{\rm QSO} \rangle = \phi_{\rm QSO}^{-1/3}/(1+z) \approx 30h_{70}^{-1}$ 
Mpc.  As discussed by Giroux, Fardal, \& Shull (1995), the ``proximity 
spheres" for bright QSOs are $\sim$1 Mpc at 1 ryd but as high as 35 Mpc at 
4 ryd.  At these distances, the QSOs's local radiation field matches the 
metagalactic background and can create detectable changes in the H~I and 
He~II line opacity. In the case of HE~2347-4342, Smette et al.\ (2002) 
noted the absence of any observed proximity effect in He~II absorbers near 
$z_{\rm em} = 2.885$.  They estimated that a column density 
N(He~II) $\geq 2.4 \times 10^{18}$ cm$^{-2}$ was required to absorb the QSO's 
ionizing radiation at 4 ryd.  Thus, the QSO radiation field in the He~II 
continuum may be strongly affected by absorption from both the galactic 
nucleus and the IGM. 

If a significant fraction of AGN exhibit He~II intrinsic absorption, then 
the ratio of fluxes at 1 and 4 ryd can be larger than implied by the range
in $\alpha_{\rm EUV}$ (Telfer et al.\ 2002), which is only fitted between
1200 and 500~\AA.  Moreover, if this intrinsic absorption partially 
obscures the nuclear continuum in only some directions, other QSOs off the 
sightline could create regions of hard and soft ionizing radiation.  
These AGN could produce fine-grained variations in $\eta$, owing to 
their own patchy absorption.  This idea was proposed for a 
``helium-reionizing quasar" 3.2 Mpc off the sightline to Q0302-003
(Jakobsen et al.\ 2003).  
 
We now can characterize the difficulty with the simple explanation of 
attributing fluctuations in $\eta$ to variable QSO spectral indices.
The FUSE/VLT data suggest that the local radiation field varies on 1~Mpc 
scales, whereas bright QSOs are 30 Mpc apart, on average.  However, 
their proximity spheres extend to 1 Mpc at 1 ryd, and to 35 Mpc at 4 ryd. 
The observed fluctuations in $\eta$ are produced by the combination of
the wide range in source emissivities and AGN/IGM opacities at 1 and 4 ryd.  
The non-uniformity of the metagalactic background, $J_{\nu}$, needs to be 
confirmed by numerical simulations that carefully treat the radiative 
transfer (absorption, filtering, reprocessing) of ionizing radiation through 
the cosmic web and include the range of QSO spectral indices.
In standard models of the ionizing background (FGS98; Madau et al.\ 1999), 
the radiation field at a given point is produced by many ionizing sources 
within a sphere of radius $r_0$, the comoving attenuation length for 
``optical depth unity" in the continuum.  Fluctuations in the ionizing 
background are then set by statistics of sources within that sphere.  

Fardal \& Shull (1993) estimated that $N_{\rm QSO} = 300-1000$ QSOs lie 
within this ``attenuation sphere" at $z \approx 2.5$ 
for a standard QSO luminosity function (Boyle et al.\ 1991).  From these
large numbers, one naively expects the H~I ionizing background to be fairly 
uniform.  However, these estimates differ at 1 ryd and 4 ryd.  FGS98 found 
that, at $z \approx 2.5$, the H~I and He~II continuum opacity gradients,  
$d \tau_{\rm cont} /dz$ are 2.5 (H~I) and 10 (He~II). 
Therefore, the spheres of optical depth unity extend over 
$\Delta z \approx 0.4$ (H~I physical length $140 \pm 10$ Mpc) 
and $\Delta z \approx 0.1$ (He~II physical length $34 \pm 1$ Mpc).

These results predict much larger intensity fluctuations 
in the He~II continuum than in the H~I continuum, both because the He~II 
attenuation sphere is smaller, and because the flat-spectrum QSOs 
($\alpha_s = 0-1.5$) are rarer than the the steep-spectrum QSOs 
($\alpha_s = 1.5-3.0$), an effect seen clearly in Figure 4 and in
Telfer et al.\ (2002). In numerical hydrodynamic simulations, 
the IGM is distributed non-uniformly in filaments, and the spherical 
approximations break down. The propagation of 1 ryd and 4 ryd radiation 
differ considerably, owing to the $\nu^{-3}$ frequency dependence of the 
opacity.  More sophisticated models should be able to confirm our hypothesis 
that the observed $\eta$-fluctuations are the result of non-uniformities in 
opacity, coupled with the observed range in QSO spectral indices.

\section{SUMMARY OF SCIENTIFIC IMPLICATIONS}

This paper continued the study of the He~II/H~I absorption toward
HE~2347-4342, using both FUSE and VLT spectra to derive values of
$\eta = 4 \tau_{\rm HeII}/\tau_{\rm HI}$ ranging from 0.1 to 460.   
The combination of He~II and H~I absorption data allows us to probe
the metagalactic ionizing spectrum and its sources.  The large observed
fluctuations in the H~I and He~II Ly$\alpha$ absorption, and the wide
range (0.1--460) in the ratio $\eta \equiv$ N(He~II)/N(H~I) suggest a
wide range of ionizing sources and filtering of the ionizing spectra by
a filamentary IGM.  The fine-grained variations in $\eta$ indicate that
the ionizing background is far from uniform, probably due to variations
in both source emissivity (spectral index) and opacity fluctuations.
A recent HST survey (Telfer et al.\ 2002) of QSO ionizing spectra
suggests that most of the observed variations ($20 < \eta < 200$)
can be understood through the wide range in QSO spectral indices
($0 < \alpha_s < 3$).

\noindent
The most significant scientific results of our study are:
\begin{itemize}

\item The observed fine-scaled ($\Delta z = 10^{-3}$) variations in $\eta$,
      on physical scales $\sim 1.3 h_{70}^{-1}$ Mpc are much smaller than
      typical distances between AGN or Lyman-limit absorbers. 

\item The wide range in $\eta$ (from 0.1--460) implies a broad 
      distribution of effective ionizing spectral index, $\alpha_s$, 
      ranging between $-1$ and 4 ($F_{\nu} \propto \nu^{-\alpha_s}$). 
      This range exceeds the observed range of AGN spectral indices,
      suggesting additional spectral softening due to radiative 
      transfer through the AGN and a filamentary cosmic web. 

\item An observed systematic correlation of higher $\eta$ (softer 
      radiation fields) for absorbers in voids in the H~I distribution 
      ($\tau_{\rm HI} < 0.05$).     

\item An inferred distribution in He~II line optical depth,
      N($\tau_{\rm HeII}) \propto \tau_{\rm HeII}^{-1}$ between
      $0 < \tau_{\rm HeII} < 1$.  

\end{itemize}

We conclude with a discussion of future needs and challenges.
The data on HE~2347-4342 ($z_{\rm em} = 2.885$) provide important 
evidence on the epoch of He~II reionization ($z_r = 2.8 \pm 0.2$), and
on the spectral variations and spatial distribution of ionizing sources.   
Nevertheless, one always worries whether cosmic variance might produce
differences along other sightlines.
It would be valuable to obtain He~II absorption data along several other 
FUSE or HST sightlines, over the critical redshift range $2.4 < z < 3.2$.  
At least one other sightline (HS~1700+6416 at $z_{\rm em} = 2.72$) has just 
been observed with FUSE, and the results look promising.   

One of the major theoretical challenges is to understand why the He~II 
reionization epoch ($z_{\rm HeII} = 2.8 \pm 0.2$) occurs much later 
than the H~I reionization epoch ($z_{\rm HI} = 6.2 \pm 0.2$). 
Did He~II experience ``double re-ionization", after an initial 
ionization epoch at $z = 6-10$ by Pop~III, low-metallicity stars 
(Venkatesan, Tumlinson, \& Shull 2003)?  Vestiges of this effect 
might still be visible in low-density voids, although much of the
effect will be washed out by the rapid increase in QSO fluxes at
$z < 6$.  The observed high $\eta$ that we
find in the voids would seem to rule out local hard spectra and 
relic ionization from early metal-free star formation.

Our finding that $\eta$ varies on small (1 Mpc) distance scales poses a
quandary for the nature of the ionizing sources.  Although it is
tempting to conclude (KS01) that the wide range of AGN spectral indices
can explain much of the He~II/H~I variation ($20 < \eta < 200$), one
still must confirm, through numerical simulations of ionizing sources
embedded in the cosmic web, that these $\eta$-variations actually 
occur naturally. 

Another intriguing result from our study is that $\eta$ is systematically 
larger in voids ($\tau_{\rm HI} < 0.05$), compared 
to filaments of H~I Ly$\alpha$ absorbers.  Does a higher $\eta$ really
imply a softer ionizing radiation field, characteristic perhaps of
starburst galaxies? Can we therefore use He~II absorbers to probe the 
distribution of gas in voids?

If the variations $\eta$ can be explained by local fluctuations in the
ionizing radiation field, enhanced by radiation transfer through the IGM,
there may be additional signatures that take place around the epoch of
He~II recombination, $z_{\rm HeII} = 2.8 \pm 0.2$.  It is still not certain
that the effect is entirely the result of variable-index QSOs;
soft spectra can also arise from starburst galaxies.
Deep imaging of these fields at $2.4 < z < 2.8$ might detect the
actual sources of ionizing radiation. At $z \approx 2.5$, an $L^*$ galaxy
($m_B = -19.5$, with 21 Gpc luminosity distance) would have magnitude 
$m_B \approx 27$.  One would need 
to search angular scales out to $2'$ from the sightline (1 Mpc offset at 
$z = 2.5$).  One might also hope to obtain additional physical information 
on the ionization state of these H~I and He~II absorbers by looking for the
corresponding C~IV, Si~IV, and O~VI absorption in the NUV and optical.

\acknowledgements 

Financial support for the FUSE work was provided by NASA contract NAS5-32985
from Johns Hopkins University and from NASA grant NAG5-13013 at the University
of Colorado. J.M.S. and M.L.G. acknowledge support at the Colorado astrophysical 
theory program from NASA LTSA grant NAG5-7262 and NSF grant AST02-06042. 
\clearpage

\begin{deluxetable}{ccc}
\tablecolumns{10}
\tablenum{1}
\tablewidth{0pt}
\tablecaption{Background Extraction Windows$^a$}
\tablehead{
\colhead{Segment}
&\colhead{Extraction Window}
&\colhead{Background Windows}
}
\startdata
 LiF1a &   530 - 609  & 450 - 499, 720 - 769, 860 - 909   \\
 LiF1b &   470 - 607  & 300 - 449, 650 - 799              \\
 LiF2a &   660 - 709  & 600 - 659                         \\
 LiF2b &   690 - 729  & 600 - 636, 734 - 770              \\
\enddata
\tablenotetext{a}{We sum data from all columns ($x$) between the listed
     columns ($y$ values above).}
\end{deluxetable}

\clearpage

\begin{deluxetable}{lcccc}
\tablecolumns{10}
\tablenum{2}
\tablewidth{0pt}
\tablecaption{Measured Transmission in He~II}
\tablehead{
\colhead{Band (\AA)}
&\colhead{Kriss et al. (2001)}
&\colhead{Smette et al. (2002)}
&\colhead{This work}
}
\startdata
1157.75-1158.25 & 0.69 $\pm$ 0.06  & 0.78 $\pm$ 0.09 & 0.80 $\pm$ 0.07  \nl
1159.0-1161.0   & 0.68 $\pm$ 0.03  & 0.64 $\pm$ 0.04 & 0.63 $\pm$ 0.04  \nl
1161.5-1168.3   & 0.02 $\pm$ 0.02  & 0.06 $\pm$ 0.02 & 0.04 $\pm$ 0.02  \nl
1174.0-1175.0   & 1.03 $\pm$ 0.06  & 0.93 $\pm$ 0.04 & 0.94 $\pm$ 0.06  \nl
1175.1-1179.3   & 0.07 $\pm$ 0.03  & 0.02 $\pm$ 0.02 & 0.00 $\pm$ 0.03
\enddata
\end{deluxetable}

\begin{figure*}[t]
\centerline{\epsfxsize=0.8\hsize{\epsfbox{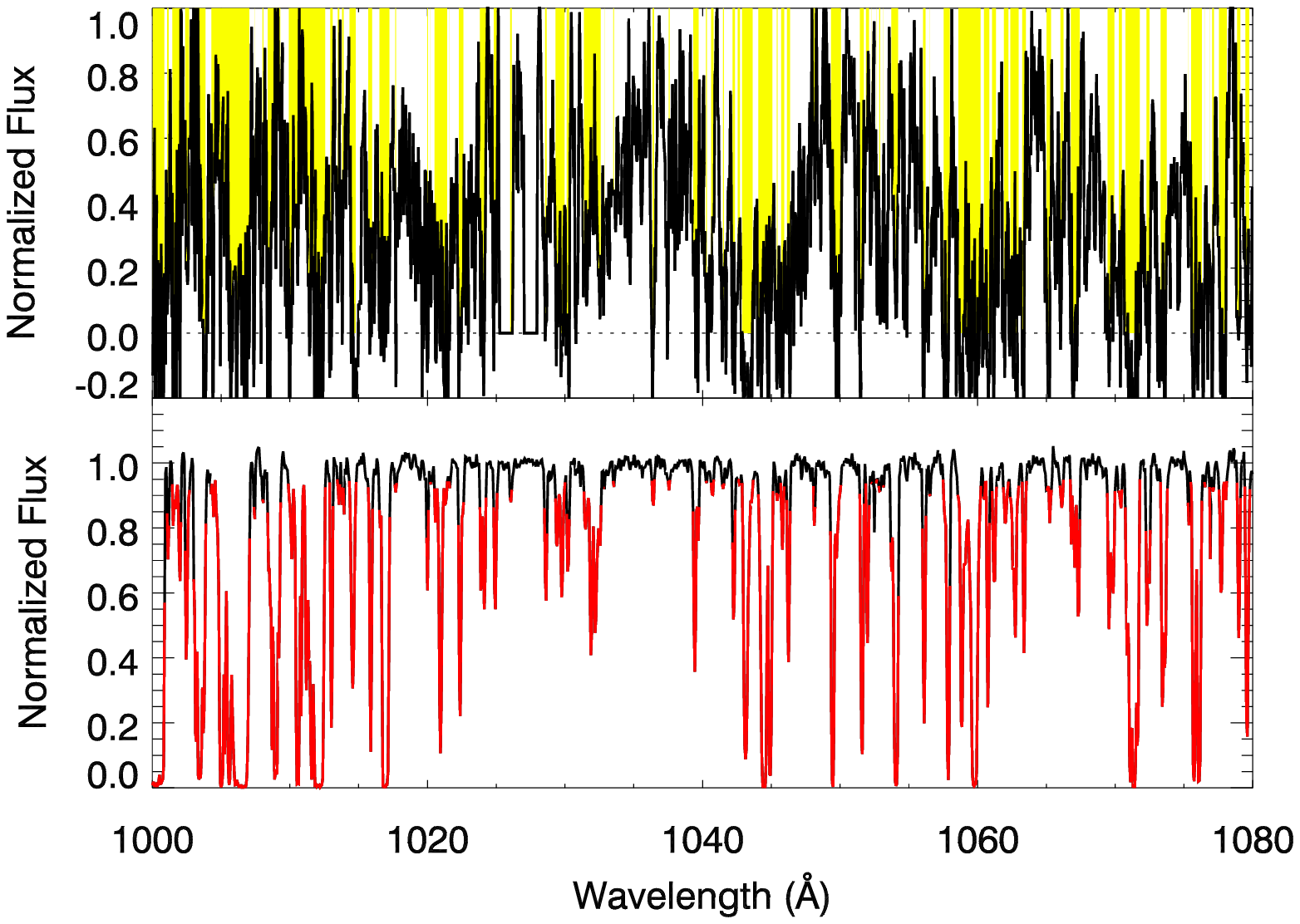}}}
\centerline{\epsfxsize=0.8\hsize{\epsfbox{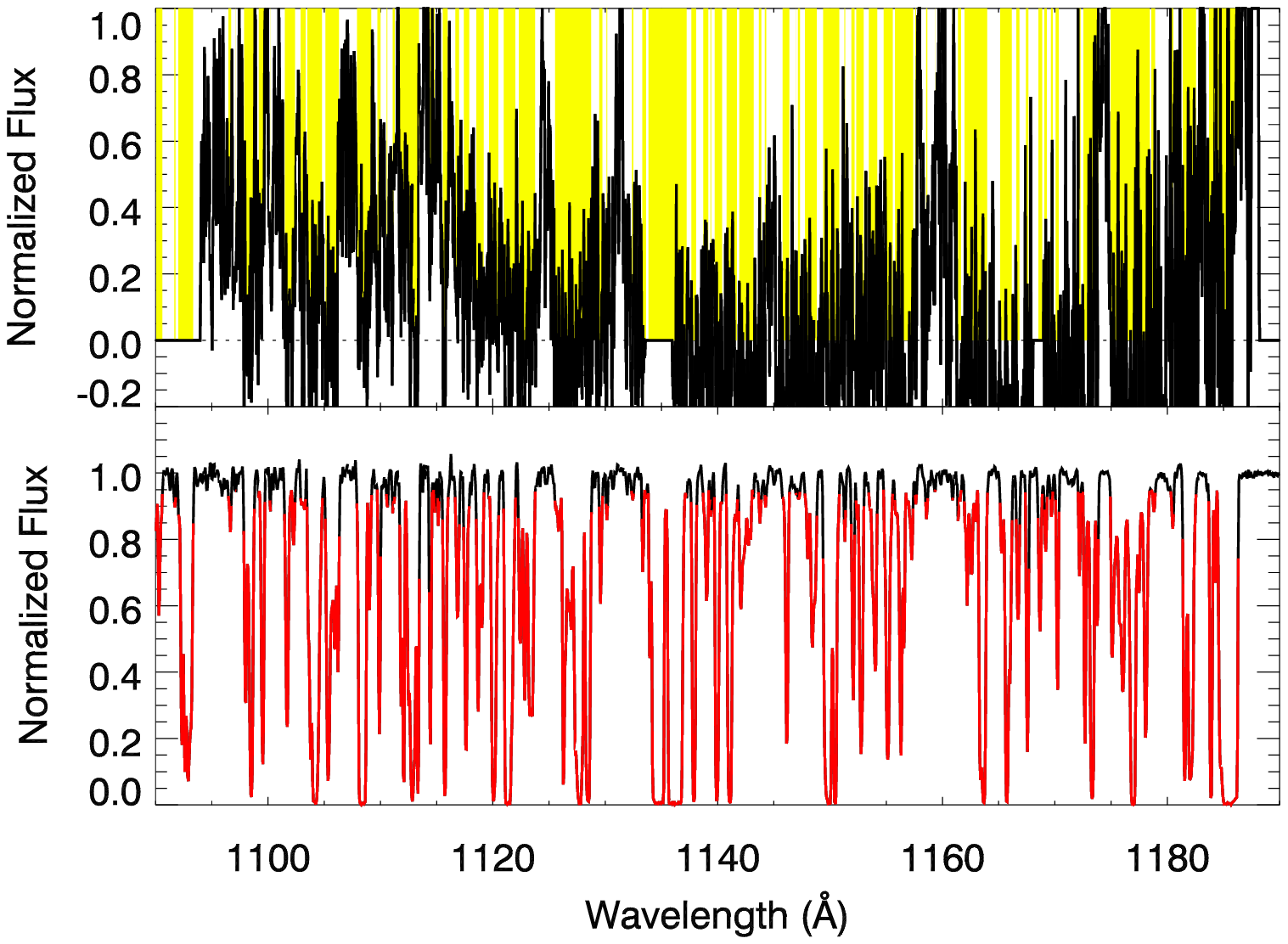}}}
\vspace{0.2in}
\figcaption{Overlay of the Ly$\alpha$ absorption of He~II (from FUSE --
top) and of H~I (from VLT/UVES -- bottom).  Wavelengths for the lower 
panels (H~I data) are divided by 4.00178 to align them with the He~II data.
The extrapolated STIS continuum is overlaid for the FUSE data.  In the 
VLT data, ``filament" points with $\tau_{\rm HI} > 0.05$ are shown in red 
and blocked in yellow above.
\label{specfig}}
\end{figure*}

\begin{figure*}[t]
\centerline{\epsfxsize=0.8\hsize{\epsfbox{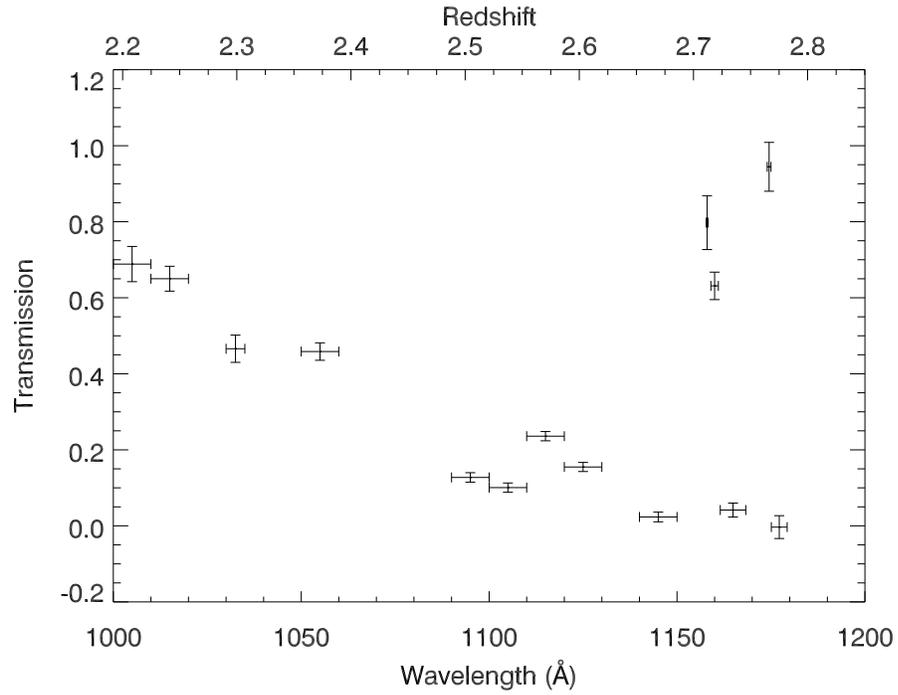}}}
\figcaption{Flux transmission through FUSE He~II absorption in several 
defined windows.  Note the three windows of high transmission between 
$z = 2.7-2.8$, flanked by low-transmission. The flux has a slow recovery of
transmission at $z < 2.4$.  Table 2 gives a comparison with previous studies.
\label{transmission} }
\end{figure*}

\begin{figure*}[t]
\centerline{\epsfxsize=1.0\hsize{\epsfbox{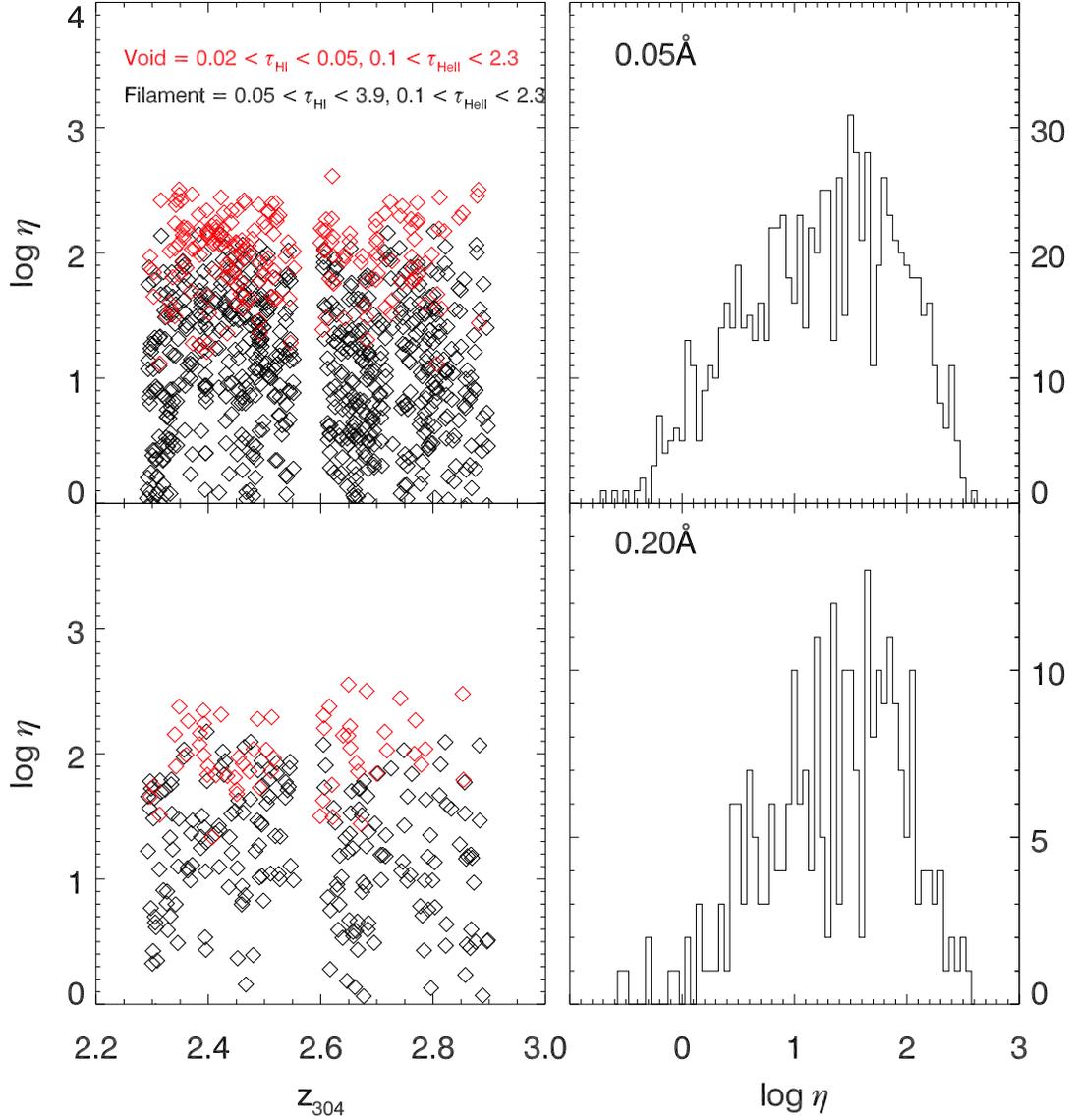}}}
\vspace{0.2in}
\figcaption{Left panels show the distribution with redshift of 
$\eta$ = N(He II)/N(H I), determined here as 
$\eta = 4 \tau_{\rm HeII}/\tau_{\rm HI}$, in wavelength bins of 
0.05~\AA\ (top panels) and 0.20~\AA\ (bottom panels).  
The distribution of $\eta$, integrated over $2.3 < z < 2.9$ is shown in
the right panels.   With the accuracy of these data, we can reliably
measure optical depths in the ranges $0.1 < \tau_{\rm HeII} < 2.3$
and $0.02 < \tau_{\rm HI} < 3.91$. Gap at $z = 2.6$ lies between the 
LiFa and LiF1b FUSE detector segments. ``Filaments'' in the Ly$\alpha$
forest are plotted in black ($\tau_{\rm HI} > 0.05$), while  ``voids'' have 
$0.02 < \tau_{\rm HI} < 0.05$ and are plotted in red, some as lower limits. 
The large fluctuations in $\eta$ suggest wide variation in the spectra of 
the ionizing sources, density fluctuations, or significant effects of 
radiative transfer in the IGM.
\label{etafig}}
\end{figure*}

\begin{figure*}[t]
\centerline{\epsfxsize=1.0\hsize{\epsfbox{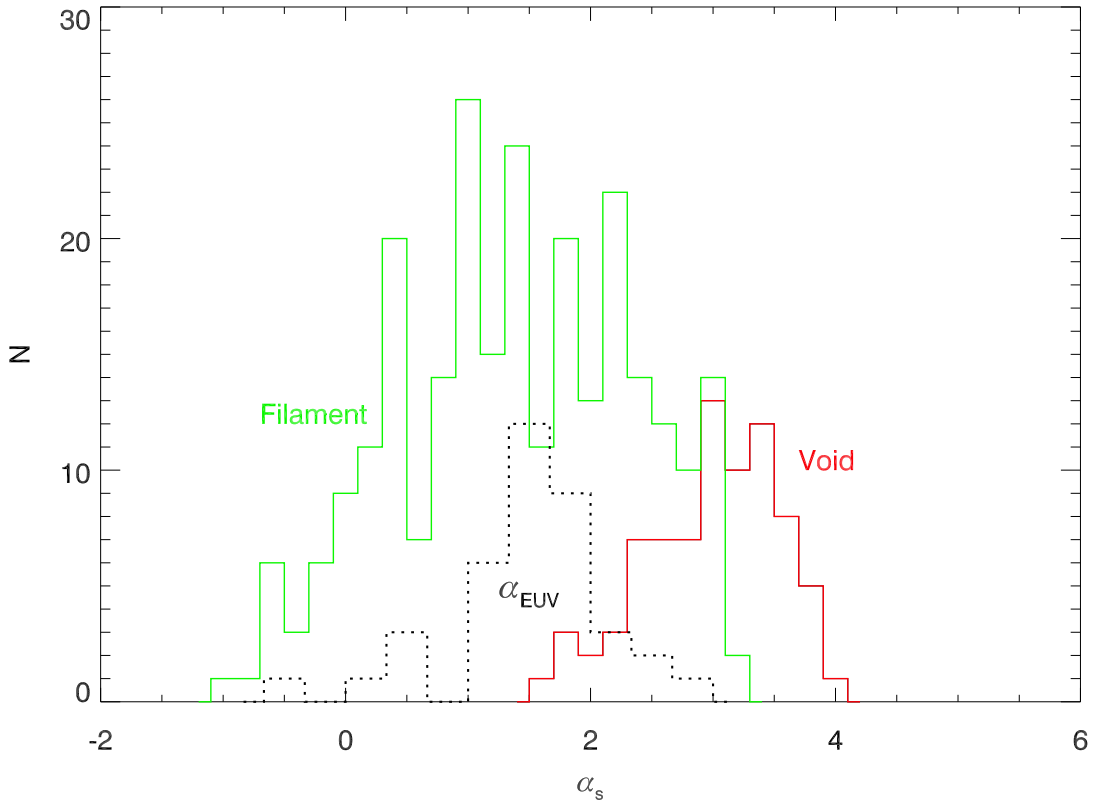}}}
\vspace{0.2in}
\figcaption{
(Bottom curve) Observed distribution of the observed QSO ionizing spectral 
indices, $\alpha_{\rm EUV}$, for radio-quiet AGN (Telfer et al.\ 2002).
The other curves show the spectral indices, $\alpha_s$, required 
(using eq. [1]) to reproduce the observed values of He~II/H~I ratios, 
Note the systematically softer radiation fields (high $\eta$, high
$\alpha_s$) in the voids (Red: $\tau_{\rm HI} < 0.05$) compared to filaments 
(Green: $\tau_{\rm HI} > 0.05$). The breadth of these curves and the
clear offset between voids and filaments suggest that additional spectral 
softening may be present.
\label{qso-alpha-dist}}
\end{figure*}

\begin{figure*}[t]
\centerline{\epsfxsize=1.0\hsize{\epsfbox{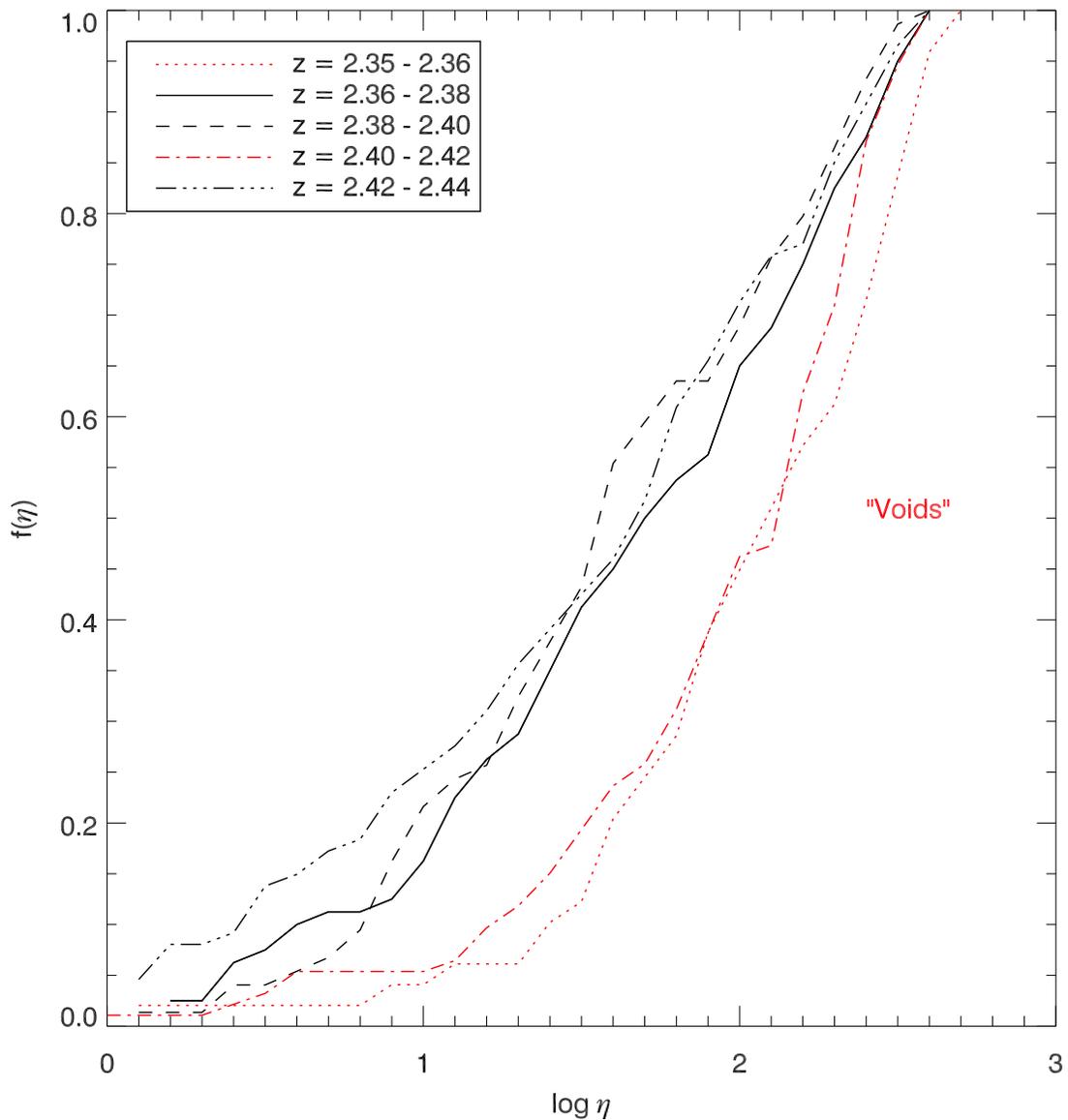}}}
\vspace{0.2in}
\figcaption{
Cumulative distribution functions of $\log \eta$. For regions with
significant H~I opacity, roughly 40\% of the bins have
$\log \eta > 2.0$. For the void region, 80\% of the bins have
$2.0 < \log \eta < 3.0$. Uniformly high $\eta$ suggests the influence
of local soft sources (e.g., widely distributed dwarf galaxies) or
heavy filtering by the IGM.  
\label{etadist}}
\end{figure*}

\begin{figure*}[t]
\centerline{\epsfxsize=0.8\hsize{\epsfbox{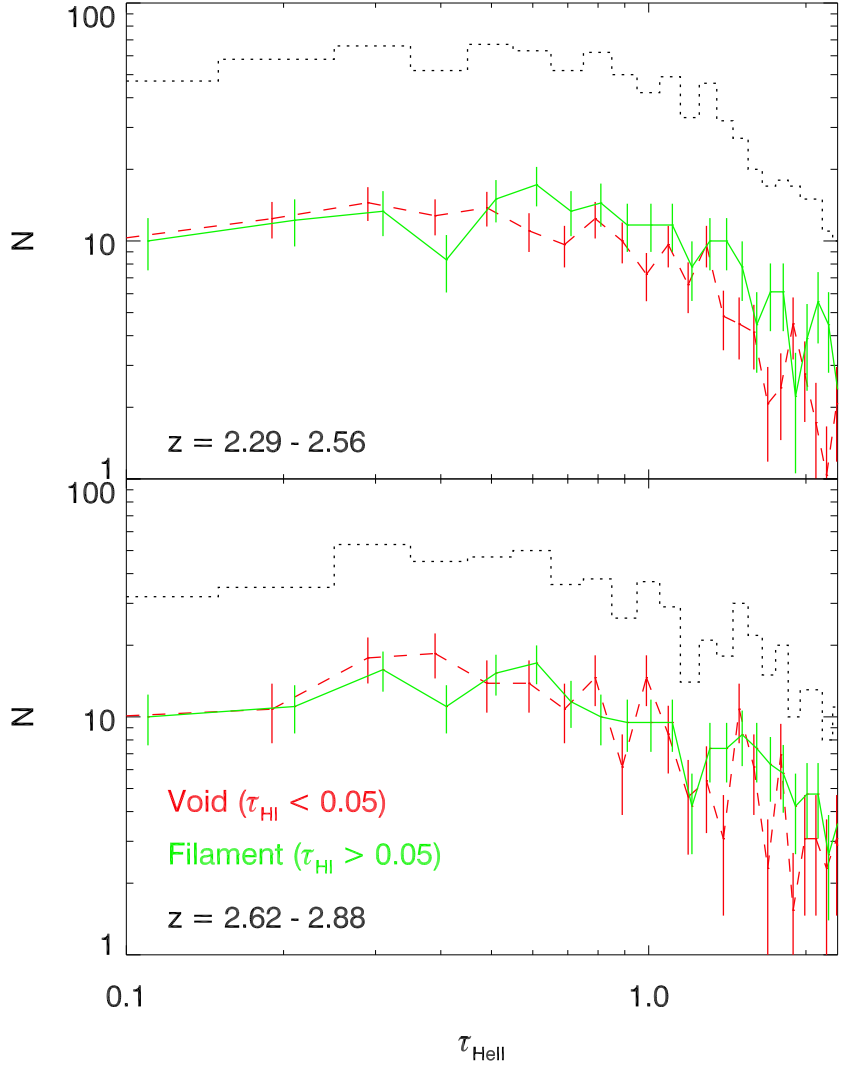}}}
\vspace{0.2in}
\figcaption{
   Distribution of optical depths, $\tau_{\rm HeII}$,   
   in two redshift intervals, $2.29 < z < 2.56$ (regions of lower optical 
   depth) and $2.62 < z < 2.88$ (regions of patchy, variable opacity.  In 
   each panel, we show distributions for ``voids" and ``filaments" in the 
   H~I Ly$\alpha$ forest. The distributions look flat (per $\log 
   \tau_{\rm HeII}$, between $0.1 < \tau_{\rm HeII} < 1$, and steepen at 
   $\tau_{\rm HeII} > 1$.  This corresponds to $N(\tau) \propto \tau^{-1}$. 
\label{taudist}}
\end{figure*}

\begin{figure*}[t]
\centerline{\epsfxsize=1.0\hsize{\epsfbox{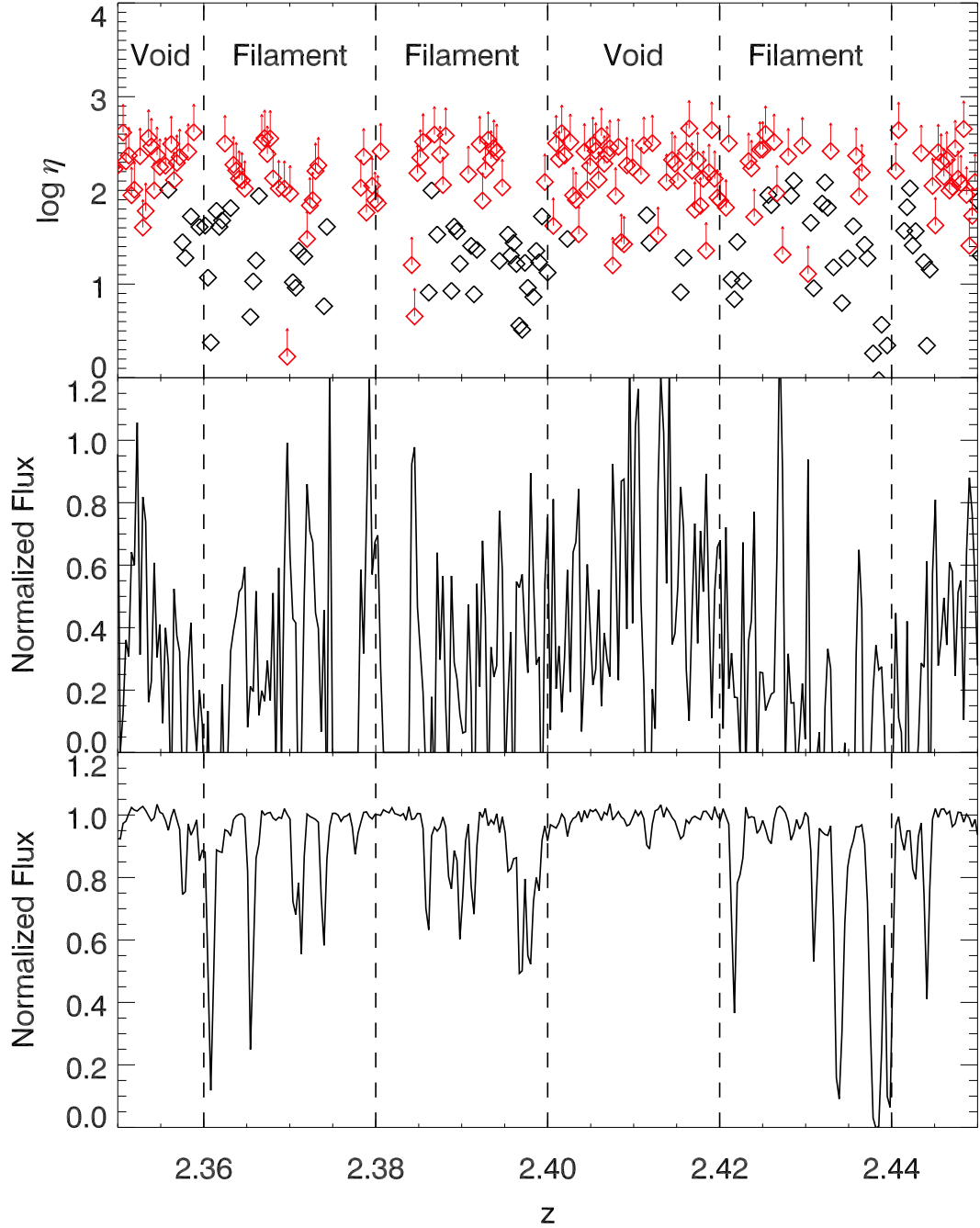}}}
\vspace{0.2in}
\figcaption{Closeup of filaments and voids ($z$ = 2.36 - 2.44). In the voids,
   ($\tau_{\rm H I} \leq 0.05$), the He~II transmission is large and  
   $\eta$ is uniformly high, perhaps indicating the influence 
   of local soft sources.
\label{voidfig}}
\end{figure*}

\end{document}